\documentclass[journal]{IEEEtran}
\usepackage{amssymb}
\usepackage{amsmath}
\usepackage{graphicx}
\usepackage{cite}
\usepackage{amsfonts}
\usepackage{booktabs}
\usepackage{multirow}
\usepackage{flushend}
\usepackage{mathrsfs}
\usepackage{comment}
\usepackage{hyperref}
\usepackage{cleveref}

\usepackage{longtable, booktabs}
\usepackage{supertabular}
\usepackage{lscape}
\usepackage{verbatim}
\usepackage{soul, color, xcolor}
\usepackage{algorithmic, algorithm}

\ifCLASSINFOpdf
\else
\fi

\begin{document}

\title{Movable Subarray-Aided ISAC in Hybrid Near-Far Field Channels}

\author{
Ruiqi Liu, Yuanshuo Gang, Honghao Wang, Tianqi Mao
\vspace{-0.5cm}

\thanks{
R. Liu is with Wireless and Computing Research Institute, ZTE Corporation, Beijing, China (e-mail: richie.leo@zte.com.cn).
Y. Gang and T. Mao are with State Key Laboratory of Environment Characteristics and Effects for Near-space, Beijing Institute of Technology, Beijing 100081, China (e-mails: \{gangys, maotq\}@bit.edu.cn). 
H. Wang is with Department of Electronic Engineering, Shanghai Jiao Tong University, Shanghai, China (e-mail: hhwang@sjtu.edu.cn).
\emph{(Corresponding authors: Yuanshuo Gang).}}
}

\maketitle
\begin{abstract}
This letter investigates an integrated sensing and communication (ISAC) system aided by movable subarrays (MSAs) using a hybrid near-far field channel model. The sensing target and communication users are assumed to lie in the near field of the overall MSA aperture but in the far-field region of each subarray. Accordingly, a hybrid near-far field channel model is established, and the equivalent Fisher information matrix and Cram\'er-Rao bound (CRB) for joint range, elevation, and azimuth estimation are derived. The transmit beamforming matrix and subarray positions are jointly optimized to minimize the trace CRB subject to minimum communication signal-to-interference-plus-noise ratio (SINR), maximum transmit power and subarray movement constraints. An alternating optimization algorithm is developed combining iterative rank-one-penalized semidefinite relaxation with projected finite-difference block descent and backtracking. Numerical results show that the hybrid-field model closely matches the spherical-wave model, while MSAs substantially reduce the CRB. 
\end{abstract}

\begin{IEEEkeywords}
Integrated sensing and communication, movable antenna, movable subarray, near-field, 6G.
\end{IEEEkeywords}

\IEEEpeerreviewmaketitle

\section{Introduction}
As the industry starts to define the sixth generation (6G) wireless system, integrated sensing and communication (ISAC) has been agreed widely as one of the promising applications \cite{10904090}. 
ISAC allows for the network to provide communication data as well as sensing services using shared resources \cite{9737357,10803090}, which creates new markets for operators and exciting opportunities for vertical industries to digitalize. 

To provide high angular resolution for sensing, a large array aperture is needed. Traditionally, this is supported by a large antenna array, which, however, induces high hardware cost and high power consumption. To harness the spatial flexibility of wireless and sensing systems, movable antenna (MA) is proposed \cite{10643473,WangHH_low_altitude_Mag}. Compared to conventional arrays with fixed antenna positions, the new degree of freedom (DoF) in MA position has a potential to significantly improve the sensing performance using a smaller number of antennas benefiting from the increased effective array aperture and joint optimization with beamforming. In \cite{10643473}, authors study both one-dimensional and two-dimensional MAs and optimize the sensing performance characterized by Cram\'er-Rao bound (CRB) under the far-field condition. 
The far-field channel model is also adopted in \cite{10839251,11436123} for linear/planar MA arrays while jointly optimizing over the antenna position and beamforming matrix.
There are also preliminary studies on MA‑aided near‑field communication and sensing systems \cite{11075964,10909572}. In an movable subarray (MSA) architecture, the users, targets, and scatterers can be in the near field of the overall aperture but in the far field of each compact subarray. This two‑scale geometry yields a separable spherical‑planar response: only the subarray‑center phase vector changes with MSA movement, while the intra‑subarray steering vector is reusable. The resulting model retains the range‑dependent inter‑subarray phase curvature and provides a lower‑dimensional position‑dependent representation suitable for subarray‑wise channel reconstruction

In light of the above, this letter investigates an MSA‑aided monostatic ISAC system under the hybrid propagation regime. We establish a hybrid near-far field model that retains the spherical wave phase variation across MSA centers while reusing the planar intra‑subarray response. Based on this model, the equivalent Fisher information matrix (FIM) and CRB for joint range, elevation and azimuth estimation are derived, and the transmit beamforming matrix and MSA positions are jointly optimized. An alternating optimization (AO) algorithm combining rank‑one‑penalized SDR and projected finite‑difference position descent is developed. Numerical results verify the small mismatch relative to the element‑wise spherical‑wave model and demonstrate the sensing gain enabled by MSA mobility.

\textit{Notations:} $\mathbf{A}^T$, $\mathbf{A}^H$, $\operatorname{tr}(\mathbf{A})$ and $\operatorname{vec}(\mathbf{A})$ denote the transpose, conjugate transpose, trace and vectorization of matrix $\mathbf{A}$. $[\mathbf a]_n$ denotes the $n$-th element of vector $\mathbf a$.  $\otimes$ denotes the Kronecker product. $\mathbf I_N$ denotes the $N \times N$ identify matrix.
\vspace{-0.1cm}

\section{System and Channel Model} \label{Sec:system_model}
\begin{figure}[t]
    \centering
    \includegraphics[width=0.43\textwidth]{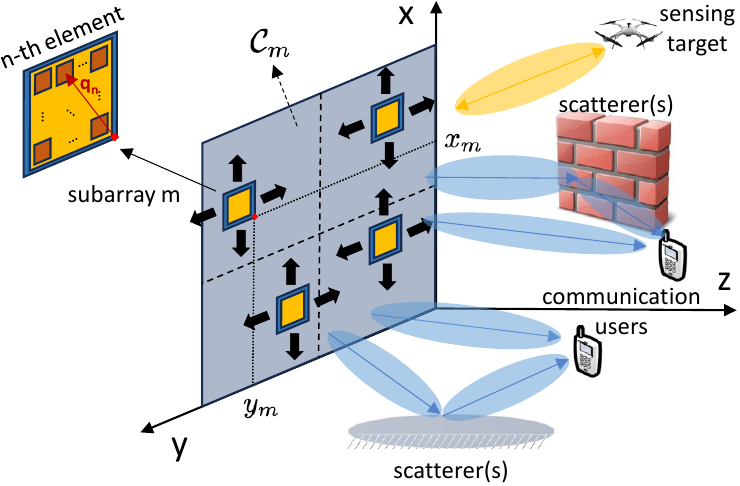}
    \caption{System model of the MA-aided ISAC system.}
    \label{fig:system_model}
    \vspace{-5 mm}
\end{figure}

In this letter, a mono-static ISAC system utilizing a BS with MSAs is considered. The study focuses on a target-tracking mode where a predicted or previous target state is available. The BS serves $K$ communication users which all utilize a single antenna. The BS also performs sensing task to sense a single target which is modeled as a point scatterer. The BS is equipped with $M$ MSAs which are uniform planar arrays (UPAs), each comprising of $N = N_x N_y$ antenna elements with half-wavelength spacing. Thus, the total number of antennas at the BS is $MN$. The BS works in time division duplexing mode and subarrays remain static during a same round of transmission and reception. As depicted in Fig. \ref{fig:system_model}, the reference point of the $m$-th MSA is $\mathbf t_m=[x_m,y_m,0]^T \in\mathcal C_m $, where $\mathcal C_m$ represents the square area for the $m$-th subarray to move within. Further, denote the location offset of the center of the $n$-th antenna element to the reference point of the $m$-th subarray as $\mathbf q_n$ so that the coordinates of the $n$-th antenna element of the $m$-th subarray are expressed as $\mathbf t_{m,n}=\mathbf t_m+\mathbf q_n$. Additionally, the reference points of all subarrays are denoted as $\mathbf T\triangleq\{\mathbf t_1,\ldots,\mathbf t_M\}$. Later in the letter, for the sake of brevity, the position of a subarray refers to the position of the reference point of the subarray and the position of an individual antenna element refers to the center point of that element.

For the $k$-th user, define the set of propagation paths from the BS to the user as $\mathcal L_k\triangleq\{0,1,\ldots,L_k\}$, where $\ell=0$ represents the line-of-sight (LOS) path and $\ell\ge 1$ represents the non-LOS (NLOS) path. The positions of the scatterers are denoted by $\mathbf r_{k,\ell}\in\mathbb R^3$, corresponding to the scatterer generating the $\ell$-th path of user $k$. $\mathbf r_{k,0}$ denotes the position of the $k$-th communication user, corresponding to the LOS case.

In this letter, a hybrid far-field and near-field scenario is considered. Particularly, the sensing target, communication users and scatterers are within the far-field region for each MSA but are in the near field for the whole BS antenna array. 
As a result, the spherical-wave propagation needs to be considered among subarrays, since the whole array aperture is large enough to create near-field effect. Comparatively, the aperture of each subarray is much smaller, and therefore, the channel responses within each subarray follow the planar-wave model.
To model the hybrid near- and far-field channel, the distance from $m$-th subarray to $\mathbf r_{k,\ell}$ is denoted as $d_m^{k,\ell} = \left\|
\mathbf r_{k,\ell}-\mathbf t_m \right\|_2 $.
Among the subarrarys, the channel response is expressed according to the spherical model as\vspace{-0.1cm} 
\begin{equation} \label{eq:comm_spherical}
\begin{split}
[\mathbf b_{k,\ell}(\mathbf T)]_m = \exp( -j\frac{2\pi}{\lambda} d_m^{k,\ell} ) ,
\end{split}
\end{equation}
where $\mathbf b_{k,\ell}(\mathbf T)\in\mathbb C^{M\times 1}$ and $m=1,... ,M$. Define the propagation direction from the center point of BS antenna array to the $k$-th user through the $\ell$-th path as $\mathbf u_{k,\ell}$, then the channel elements within the subarray for path $(k,\ell)$ can be expressed using the planar-wave model as\vspace{-0.1cm} 
\begin{equation} \label{eq:comm_planar}
[\mathbf a_{k,\ell}]_n = \exp( j\frac{2\pi}{\lambda} \mathbf u_{k,\ell}^T\mathbf q_n ),
\end{equation}
where $\mathbf a_{k,\ell}\in\mathbb C^{N\times 1}$ and $n=1,...,N$. 

By combining the spherical-wave propagation among subarrays and the planar-wave propagation within each subarray \cite{9663284}, the channel of path $(k,\ell)$ can be expressed as\vspace{-0.1cm}
\begin{equation}
\mathbf g_{k,\ell}(\mathbf T) = \mathbf b_{k,\ell}(\mathbf T)\otimes \mathbf a_{k,\ell} \in\mathbb C^{MN\times 1}.
\end{equation}

Let
$\mathbf G_k(\mathbf T)
\triangleq [\mathbf g_{k,0}(\mathbf T),\ldots, \mathbf g_{k,L_k}(\mathbf T)]$.
The physical one-way propagation coefficient vector from the BS antennas to user $k$ is $\mathbf c_k(\mathbf T)=\mathbf G_k(\mathbf T)\boldsymbol{\beta}_k$. Following the standard communication-channel convention, we define the channel vector used in the Hermitian input-output
model as 
\begin{equation}\label{eq:comm_channel}
\begin{aligned}
\mathbf h_k(\mathbf T)
\triangleq\mathbf c_k^*(\mathbf T)
= \left(\mathbf G_k(\mathbf T)\boldsymbol{\beta}_k
\right)^*
= \sum_{\ell=0}^{L_k}
\beta_{k,\ell}^*\mathbf g_{k,\ell}^*(\mathbf T).
\end{aligned}
\end{equation}
Here,
$\boldsymbol{\beta}_k=[\beta_{k,0},\ldots,\beta_{k,L_k}]^T$
collects the physical complex path coefficients referenced to the BS origin. 
We adopt a uniform-power hybrid-field model, under which the subarray movement changes the propagation phases while the amplitude variation over the movable aperture is neglected. This approximation is accurate when the path range is much larger than the array radius.


The channel between the BS and the sensing target can be derived following a same methodology. Consider a sensing target $\mathbf r_s\in\mathbb R^3$, the distance between the $m$-th subarray and the target is $d_m^s = \left\| \mathbf r_s-\mathbf t_m \right\|_2$. Similar to \eqref{eq:comm_spherical}, the spherical-wave phase response from the $m$-th subarray reference point to the sensing target can be written as\vspace{-0.1cm}
\begin{equation} \label{eq:sensing_spheircal}
[\mathbf b_s(\mathbf T)]_m = \exp( -j\frac{2\pi}{\lambda} d_m^s ).
\end{equation}
Define the direction from the BS central point to the sensing target as $\mathbf u_s$. Similar to \eqref{eq:comm_planar}, the $n$-th entry of the intra-subarray planar-wave response associated with the sensing target is\vspace{-0.1cm}
\begin{equation} \label{eq:sensing_planar}
[\mathbf a_s]_n = \exp( j\frac{2\pi}{\lambda} \mathbf u_s^T\mathbf q_n).
\end{equation}

By combining \eqref{eq:sensing_spheircal} and \eqref{eq:sensing_planar}, the hybrid near- and far-field channel response for a sensing target can be expressed as\vspace{-0.1cm} 
\begin{equation} \label{eq:sense_g_s}
\mathbf g_s(\mathbf T) = \mathbf b_s(\mathbf T)\otimes \mathbf a_s \in\mathbb C^{MN\times 1}.
\end{equation}
Under mono-static sensing mode, the channel response matrix of the reflected wave from the target to the BS can be written as\vspace{-0.1cm} 
\begin{equation} \label{eq:sensing_channel}
\mathbf H_s(\mathbf T) = \beta_s \mathbf g_s(\mathbf T)\mathbf g_s^T(\mathbf T),
\end{equation}
where $\beta_s$ denotes the unknown effective complex target reflection coefficient, which reflects the target scattering characteristics, common two-way propagation attenuation, and scattering phase. Under the adopted uniform-power model, $\mathbf g_s(\mathbf T)\mathbf g_s^T(\mathbf T)$ captures the position-dependent two-way spatial phase response, so these geometric phases are not included again in $\beta_s$.
Given the above, the received sensing signal at the BS at time $i$ is \vspace{-0.1cm} 
\begin{equation} \label{eq:sense_received_sig}
\mathbf y_s[i] = \mathbf H_s(\mathbf T)\mathbf x[i]+\mathbf n_s[i],
\end{equation}
where $\mathbf x[i]$ is transmitted signal and $\mathbf n_s[i]\sim\mathcal{CN}(\mathbf 0,\sigma_s^2\mathbf I_{MN})$ is the additive white Gaussian noise (AWGN) at the receiver. 
\vspace{-0.1cm}

\section{Problem Formulation} \label{Sec:problem_formulation}
Based on the system and channel models derived in Section \ref{Sec:system_model}, we characterize the performance for the communication users as well as the sensing performance, and eventually introduce the optimization problem to find the suitable subarray positions as well as the beamforming matrix to use.

Firstly, we derive the signal-to-interference-plus-noise ratio (SINR) to represent communication performance. Express the signal transmitted by the BS at time $i$ as 
$\mathbf x[i] = \sum_{k=1}^{K}\mathbf w_k s_k[i]$, where $\mathbf w_k\in\mathbb C^{MN\times 1}$ and $s_k[i]$ denote the transmit beamforming vector and the symbol for user $k$, respectively. 
$\mathbb E\{{|s_k[i]|^2}\}=1$ and data for different users are uncorrelated.
Let $\mathbf W=[\mathbf w_1,\ldots,\mathbf w_K]$ denote the beamforming matrix used for all users. The received signal at  user $k$ is\vspace{-0.1cm}
\begin{equation}
y_k[i] = \mathbf h_k^H(\mathbf T) \mathbf x[i]+n_k[i],
\end{equation}
where $\mathbf h_k(\mathbf T)$ is defined in \eqref{eq:comm_channel} and $n_k[i]\sim\mathcal{CN}(0,\sigma_k^2)$ is the AWGN. Accordingly, the SINR is\vspace{-0.1cm} 
\begin{equation}
\gamma_k(\mathbf W,\mathbf T) = \frac{\left| \mathbf h_k^H(\mathbf T)\mathbf w_k \right|^2}
{\sum_{j\ne k} \left| \mathbf h_k^H(\mathbf T)\mathbf w_j \right|^2 + \sigma_k^2}.
\end{equation}

Secondly, we derive the CRB according to the channel model for sensing. Let $\mathbf r_s=r_s\mathbf u_s(\theta_s,\phi_s)$ denote the position of the sensing target, where $r_s$, $\theta_s$ and $\phi_s$ represent its range, elevation and azimuth and $\mathbf u_s(\theta_s,\phi_s)$ is the direction vector for the target.
Following \eqref{eq:sense_g_s}, denote 
\begin{equation}  \label{eq:G_g_g}
\mathbf G_s(\boldsymbol{\eta},\mathbf T)
= \mathbf g_s(\mathbf T)\mathbf g_s^T(\mathbf T),
\end{equation}
where $\boldsymbol{\eta}=[r_s,\theta_s,\phi_s]^T$.  As a common operation in ISAC systems, $T_s$ snapshots are collected. Denote $\mathbf X=[\mathbf x[1],\ldots,\mathbf x[T_s]]$. Substituting \eqref{eq:sensing_channel} and \eqref{eq:G_g_g} into \eqref{eq:sense_received_sig}, the received signal at the BS side after target reflection is
\begin{equation}
\mathbf Y_s = \beta_s\mathbf G_s(\boldsymbol{\eta},\mathbf T)\mathbf X + \mathbf N_s. 
\end{equation}

To derive the FIM, we define 
$\boldsymbol{\mu} = \operatorname{vec} ( \beta_s\mathbf G_s(\boldsymbol{\eta},\mathbf T)\mathbf X )$. 
Since $\beta_s$ is typically unknown, we define the real and imaginary part of $\beta_s$ as nuisance parameters and $\boldsymbol{\beta} = [\Re\{\beta_s\},\Im\{\beta_s\}]^T$. Further, let 
\begin{equation}
\boldsymbol{\xi} = [\boldsymbol{\eta}^T,\boldsymbol{\beta}^T]^T
= [r_s,\theta_s,\phi_s,\Re\{\beta_s\},\Im\{\beta_s\}]^T
\end{equation} 
be defined as the extended unknown parameter vector. Finally, the element in the FIM $\mathbf J_{\boldsymbol{\xi}}$ can be expressed as 
\begin{equation}
[\mathbf J_{\boldsymbol{\xi}}]_{p,q}
= \frac{2}{\sigma_s^2} \Re\left\{ ( \frac{\partial \boldsymbol{\mu}} {\partial \xi_p}
)^H ( \frac{\partial \boldsymbol{\mu}}
{\partial \xi_q} ) \right\}. 
\end{equation} 

Let $\mathbf R_x=\frac{1}{T_s}\mathbf X\mathbf X^H$ denote the co-variance matrix of the transmitted signal. For the sake of brevity, 
$\mathbf G_{s,p} \triangleq \frac{\partial \mathbf G_s}{\partial \eta_p}, \quad \eta_p\in\{r_s,\theta_s,\phi_s\}$,  $\Omega_{p,q}
\triangleq \operatorname{tr}
( \mathbf G_{s,q}\mathbf R_x\mathbf G_{s,p}^H)$, $\zeta_p \triangleq \operatorname{tr} (\mathbf G_s\mathbf R_x\mathbf G_{s,p}^H)$ and $\nu \triangleq
\operatorname{tr} (\mathbf G_s\mathbf R_x\mathbf G_s^H)$.
In this way, the FIM can be expressed block-wise as 
\begin{equation}  \label{eq:FIM} \mathbf J_{\boldsymbol{\xi}}=\begin{bmatrix}\mathbf J_{\boldsymbol{\eta}\boldsymbol{\eta}}&\mathbf J_{\boldsymbol{\eta}\boldsymbol{\beta}}\\\mathbf J_{\boldsymbol{\beta}\boldsymbol{\eta}}&\mathbf J_{\boldsymbol{\beta}\boldsymbol{\beta}}\end{bmatrix},
\end{equation}
where
$ [\mathbf{J}_{\boldsymbol{\eta\eta}}]_{p,q}=\frac{2T_s|\beta_s|^2}{\sigma_s^2}\Re\{\Omega_{p,q}\},\quad\eta_p,\eta_q\in\{r_s,\theta_s,\phi_s\}, $
$ [\mathbf J_{\boldsymbol{\eta}\boldsymbol{\beta}}]_{p,1}
= \frac{2T_s}{\sigma_s^2} \Re\{\beta_s^*\zeta_p\}$, $[\mathbf J_{\boldsymbol{\eta}\boldsymbol{\beta}}]_{p,2}
= -\frac{2T_s}{\sigma_s^2} \Im\{\beta_s^*\zeta_p\}$ and
$\mathbf J_{\boldsymbol{\beta}\boldsymbol{\beta}}
= \frac{2T_s\nu}{\sigma_s^2}\mathbf I_2$.
Using the Schur complement, the equivalent FIM after eliminating the nuisance parameters can be derived as $\mathbf J_{\rm e}
= \mathbf J_{\boldsymbol{\eta}\boldsymbol{\eta}}
- \mathbf J_{\boldsymbol{\eta}\boldsymbol{\beta}}
\mathbf J_{\boldsymbol{\beta}\boldsymbol{\beta}}^{-1}
\mathbf J_{\boldsymbol{\beta}\boldsymbol{\eta}}$,
which can also be expressed element-wise as 
\begin{equation} [\mathbf J_{\rm e}]_{p,q}
= \frac{2T_s|\beta_s|^2}{\sigma_s^2}
( \Re\{\Omega_{p,q}\} - \frac{\Re\{\zeta_p\zeta_q^*\}}{\nu}), \eta_p,\eta_q\in\{r_s,\theta_s,\phi_s\}.
\end{equation}

Based on the above deduction, the CRB for sensing target distance estimation and two-dimensional angle estimation with respect to $\mathbf W$ and $\mathbf T$ is $\mathrm{CRB}(\mathbf W,\mathbf T)=\mathbf J_{\rm e}^{-1}(\mathbf W,\mathbf T)$. Later, $\operatorname{tr}(\mathrm{CRB}(\mathbf W,\mathbf T))$ is used as a
scalarized sensing metric that jointly accounts for range, elevation, and azimuth.

Finally, an optimization problem is formulated to minimize the CRB while maintaining a certain level of SINR for communications by choosing the optimal beamforming vectors at the BS and the positions of MSAs. The corresponding optimization problem is\vspace{-0.1cm}
\begin{subequations}\label{eq:p1}
\begin{align}
\text{(P1)}:\quad
\min_{\mathbf W,\mathbf T}\quad
& \operatorname{tr}\!\left(\mathrm{CRB}(\mathbf W,\mathbf T)\right)
\label{eq:p1_objective}\\[-0.2cm]
\text{s.t.}\quad
& \gamma_k(\mathbf W,\mathbf T)\ge \Gamma_k,
\quad k=1,\ldots,K,
\label{eq:p1_sinr_constraint}\\
& \|\mathbf W\|_F^2\le P_{\max},
\label{eq:p1_power_constraint}\\
 \mathbf t_m\in\mathcal C_m,  \mathbf t_m + & [l_x,l_y,0]^T \in\mathcal C_m, 
\quad m=1,\ldots,M,
\label{eq:p1_moving_region_constraint}\\
& \|\mathbf t_m-\mathbf t_{m'}\|_2\ge D_{\min},
\quad m\ne m',
\label{eq:p1_min_distance_constraint}
\end{align}
\end{subequations}
where $\Gamma_k$ is the minimum SINR required for user $k$, $P_{\max}$ is the maximum transmit power of the BS, $l_x, l_y$ are the width and length of each subarray, $D_{\min}$ is the minimum distance between any two subarrays due to physical restrictions.

\vspace{-0.1cm}
\section{Proposed Solution} \label{Sec:solution}
One of the challenges to solve problem \eqref{eq:p1} is that it is non-convex. On one hand, the subarray positions $\mathbf T$ can impact the communication channels and the FIM through $\mathbf b_{k,\ell}(\mathbf T)$ and $\mathbf b_s(\mathbf T)$. On the other hand, the beamforming matrix $\mathbf W$ determines the received power and interference of multiple communication users as well as the received power of the reflected wave from the sensing target. Thus, the two optimization variables $\mathbf W$ and $\mathbf T$ are coupled in both the objective function (CRB) and the constraint on SINR. To address this challenge, an AO method is used to solve the problem.


Specifically, within each AO iteration, $\mathbf W$ is updated through an iterative rank-one-penalized semidefinite relaxation (SDR) for fixed $\mathbf T$. Subsequently, $\mathbf T$ is
updated via projected finite-difference block
descent with backtracking for fixed $\mathbf W$. The two update procedures are detailed below and summarized in Algorithm~\ref{alg:A1}.

\begin{algorithm}[tbp]
\caption{The proposed AO algorithm}
\label{alg:A1}
\begin{algorithmic}[1]
\STATE \textbf{Initialization:} Initialize feasible $\mathbf T^{(0)}$, $\mathbf W^{(0)}$, system parameters, and maximum iteration number $I_{\max}$.
\STATE Set $r=0$. 
\REPEAT
\STATE Given $\mathbf T^{(r)}$, solve \eqref{eq:p2.1} to obtain $\{\mathbf Q_k^\star\}$ and $\mathbf U^\star$.
\STATE Recover $\mathbf W^{(r+1)}$ from $\{\mathbf Q_k^\star\}_{k=1}^K$.
\STATE Set
$\widetilde{\mathbf T}\leftarrow\mathbf T^{(r)}$.
\FOR{$m=1,\ldots,M$}
\STATE 
Obtain $\mathbf d_m$ according to \eqref{eq:fd_descent_direction}.
\STATE Generate $\mathbf p_m^{\rm tr}$ according to
\eqref{eq:projected_position_trial}, and reduce
$\alpha_m\leftarrow\tau\alpha_m$ until
\eqref{eq:p1_sinr_constraint},
\eqref{eq:p1_min_distance_constraint}, and
\eqref{eq:position_monotone_test} are satisfied.
\STATE Accept $\mathbf p_m^{\rm tr}$ as the $m$-th block of $\widetilde{\mathbf T}$; if no admissible $\alpha_m\ge\alpha_{\min}$ exists, retain the current block.
\ENDFOR
\STATE Set
$\mathbf T^{(r+1)}\leftarrow\widetilde{\mathbf T}$.
\STATE Set $r\leftarrow r+1$.
\UNTIL{the relative decrease of \eqref{eq:p1_objective} is below $\epsilon_{\rm AO}$ or $r=I_{\max}$.}
\STATE Set optimized solutions $\mathbf T^\star=\mathbf T^{(r)}$ and $\mathbf W^\star=\mathbf W^{(r)}$.
\end{algorithmic}
\end{algorithm}

\vspace{-0.2cm}
\subsection{Updating Beamforming Matrices Using Rank-One-Penalized SDR}
With $\mathbf T$ fixed, the subproblem to optimize over the beamforming matrix $\mathbf W$ can be written as\vspace{-0.1cm}
\begin{subequations}
\begin{align}
\text{(P2)}:\quad
\min_{\mathbf W}\quad
& \operatorname{tr}\!\left(\mathrm{CRB}(\mathbf W)\right)\\[-0.2cm] 
\text{s.t.}\quad
& \eqref{eq:p1_sinr_constraint},\eqref{eq:p1_power_constraint}.
\end{align}
\end{subequations}

This subproblem is still non-convex, mainly because the CRB matrix involves the inverse of the FIM, while the SINR constraints contain quadratic fractional terms. We first handle the objective function. Note that the complete FIM is affine with respect to the transmit covariance matrix $\mathbf R_x=\mathbf W\mathbf W^H$. Define
$\mathbf E=
\begin{bmatrix}
\mathbf I_3\\
\mathbf 0_{2\times 3}
\end{bmatrix}$.
Then, the CRB matrix can be expressed as $\mathbf E^T\mathbf J_{\boldsymbol\xi}^{-1}\mathbf E$. To avoid directly dealing with the matrix inverse, we introduce an auxiliary matrix $\mathbf U\succeq\mathbf 0$ and impose $\mathbf U\succeq\mathbf E^T\mathbf J_{\boldsymbol\xi}^{-1}\mathbf E$. When $\mathbf J_{\boldsymbol\xi}\succ\mathbf 0$, according to the Schur complement, the above constraint is equivalent to the following linear matrix inequality:
\begin{equation} \label{eq:schur_complement}
\begin{bmatrix}
\mathbf J_{\boldsymbol\xi}(\mathbf R_x) & \mathbf E \\
\mathbf E^T & \mathbf U
\end{bmatrix}
\succeq \mathbf 0.
\end{equation}
Therefore, minimizing $\operatorname{tr}\!\left(\mathrm{CRB}(\mathbf W)\right)$ can be achieved by minimizing the upper bound $\operatorname{tr}(\mathbf U)$. Next, to handle the non-convex constraints, we define the auxiliary variables
$\mathbf Q_k \triangleq \mathbf w_k\mathbf w_k^H,\quad k=1,\ldots,K$. Clearly, $\mathbf Q_k$ should satisfy $\mathbf Q_k\succeq \mathbf 0$ and $\operatorname{rank}(\mathbf Q_k)=1$, for $k=1,\ldots,K$. The total transmit covariance matrix can thus be expressed as $\mathbf R_x=\sum_{k=1}^{K} \mathbf Q_k$.
Furthermore, define $\mathbf H_k=\mathbf h_k(\mathbf T^{(r)})\mathbf h_k^H(\mathbf T^{(r)})$. Then, the received power contributed by an arbitrary beam $\mathbf w_i$ at user $k$ can be written as $\left|\mathbf h_k^H(\mathbf T^{(r)})\mathbf w_i\right|^2=\operatorname{tr}(\mathbf H_k\mathbf Q_i)$.
Accordingly, the SINR constraint of user $k$ can be equivalently rewritten as\vspace{-0.1cm}
\begin{equation} \label{eq:sinr_cons_re}
\operatorname{tr}(\mathbf H_k\mathbf Q_k)
-
\Gamma_k
\sum\nolimits_{j\ne k}
\operatorname{tr}(\mathbf H_k\mathbf Q_j)
\ge
\Gamma_k\sigma_k^2.
\end{equation}

For any $\mathbf Q_k\succeq\mathbf 0$, the rank-one condition is equivalent to $\operatorname{tr}(\mathbf Q_k)-\lambda_{\max}(\mathbf Q_k)=0$. At the $z$-th penalty iteration, the reference matrix $\mathbf Q_k^{(z)}$ is fixed, and
$\mathbf u_k^{(z)}$ denotes its unit-norm dominant eigenvector. For any candidate $\mathbf Q_k\succeq\mathbf 0$, we have $\lambda_{\max}(\mathbf Q_k) \ge (\mathbf u_k^{(z)})^H\mathbf Q_k\mathbf u_k^{(z)}$, which yields the affine upper bound $\operatorname{tr}(\mathbf Q_k)-\lambda_{\max}(\mathbf Q_k) \le \operatorname{tr}(\mathbf Q_k) -(\mathbf u_k^{(z)})^H\mathbf Q_k\mathbf u_k^{(z)}$.
Here, $\mathbf Q_k$ is the optimization variable of the current penalized SDP, and its optimizer is denoted by $\mathbf Q_k^{(z+1)}$.
By adding this rank-one penalty, the following SDR-based beamforming problem is obtained:\vspace{-0.1cm}
\begin{subequations}\label{eq:p2.1}
\begin{align}
\text{(P2.1)}: 
\min_{\{\mathbf Q_k\},\mathbf U} 
& \operatorname{tr}(\mathbf U)
+ \frac{\rho^{(z)}}{P_{\max}}
\sum_{k=1}^{K}
[\operatorname{tr}(\mathbf Q_k)
- (\mathbf u_k^{(z)})^H \mathbf Q_k \mathbf u_k^{(z)}]
\label{eq:p2_penalty_objective}\\
\text{s.t.}\quad
& \sum_{k=1}^{K}\operatorname{tr}(\mathbf Q_k)
\le P_{\max},\\
& \mathbf U\succeq\mathbf 0,\quad
\mathbf Q_k\succeq\mathbf 0,\quad
\eqref{eq:schur_complement},\quad
\eqref{eq:sinr_cons_re}.
\end{align}
\end{subequations}
The penalty weight is increased geometrically as
$\rho^{(z+1)}=\tau_\rho\rho^{(z)}$, where
$\tau_\rho>1$. The inner iterations terminate when the normalized rank gap
\begin{equation}\label{eq:rank_gap_ratio}
\epsilon_{\rm rank}^{(z+1)}
= \frac{\sum_{k=1}^{K}
[\operatorname{tr}(\mathbf Q_k^{(z+1)})
- \lambda_{\max}(\mathbf Q_k^{(z+1)})]
}{ \sum_{k=1}^{K}
\operatorname{tr}(\mathbf Q_k^{(z+1)})}
\end{equation}
is below a prescribed threshold or the maximum number of penalty iterations is reached.

After solving problem \eqref{eq:p2.1} using semidefinite programming, the optimal positive semidefinite matrices $\{\mathbf Q_k^\star\}_{k=1}^{K}$ can be obtained. If $\{\mathbf Q_k^\star\}_{k=1}^{K}$ are rank-one, the beamforming matrix can be directly recovered via principal eigenvalue decomposition. Otherwise, Gaussian randomization method in SDR is adopted to generate candidate beamformers, and the feasible candidate satisfying the communication SINR constraints and the total power constraint with the smallest original CRB objective value is selected as the updated beamforming matrix.

\subsection{Updating MSA Positions via Projected Finite-Difference Descent}
With $\mathbf W$ fixed, the subarray-position subproblem is formulated as\vspace{-0.1cm}
\begin{subequations}\label{eq:p3_position_original}
\begin{align}
\text{(P3)}:\quad
\min_{\mathbf T}\quad
& \operatorname{tr}\!\left(
\mathrm{CRB}(\mathbf W,\mathbf T)
\right)
\label{eq:p3_position_objective}\\
\text{s.t.}\quad
& \eqref{eq:p1_sinr_constraint},
  \eqref{eq:p1_moving_region_constraint},
  \eqref{eq:p1_min_distance_constraint}.
\end{align}
\end{subequations}

Since \eqref{eq:p3_position_original} remains non-convex, the subarrays are cyclically updated using a projected finite-difference block-descent method. 
Define $\mathbf p_m=[x_m,y_m]^T$ and \vspace{-0.1cm}
\begin{equation}
\mathcal J_T(\mathbf p)
\triangleq
\operatorname{tr}\!\left(
\mathrm{CRB}(\mathbf W,\mathbf T(\mathbf p))
\right),
\end{equation}
where $\mathbf p=[\mathbf p_1^T,\ldots,\mathbf p_M^T]^T$.
For the $m$-th subarray, the block gradient ${\nabla}_m\mathcal J_T$ is numerically estimated by central finite differences, with projected one-sided differences adopted at the boundary of its feasible region 
$\mathcal P_m \triangleq
\left\{
\mathbf p_m\in\mathbb R^2: [\mathbf p_m^T,0]^T\in\mathcal C_m
\right\}$.
Specifically, for $c\in\{x,y\}$, define
$\mathbf p_{m,c}^{\pm} = \Pi_{\mathcal P_m}(\mathbf p_m\pm h\mathbf e_c)$,
where $h>0$ is the finite-difference interval and
$\mathbf e_c$ is the corresponding coordinate vector. Then,\vspace{-0.1cm}
\begin{equation}\label{eq:fd_position_gradient}
\left[\widehat{\nabla}_m\mathcal J_T\right]_c
= \frac{
\mathcal J_T(\mathbf p_{m,c}^{+},\mathbf p_{-m})
- \mathcal J_T(\mathbf p_{m,c}^{-},\mathbf p_{-m})
}{[\mathbf p_{m,c}^{+}]_c-[\mathbf p_{m,c}^{-}]_c
},
\end{equation}
whenever the denominator is nonzero.

If $\|\widehat{\nabla}_m\mathcal J_T\|_2=0$, the current block is retained. Otherwise, the normalized descent direction is\vspace{-0.1cm}
\begin{equation}\label{eq:fd_descent_direction}
\mathbf d_m =
-\frac{\widehat{\nabla}_m\mathcal J_T}
{\|\widehat{\nabla}_m\mathcal J_T\|_2},
\end{equation}
and a trial position is generated as\vspace{-0.1cm}
\begin{equation}\label{eq:projected_position_trial}
\mathbf p_m^{\rm tr} =
\Pi_{\mathcal P_m}
( \mathbf p_m+\alpha_m\mathbf d_m ),
\end{equation}
where $\Pi_{\mathcal P_m}(\cdot)$ denotes the Euclidean projection onto $\mathcal P_m$. Starting from $\alpha_0$, the step size $\alpha_m$ is successively reduced until the trial position satisfies the original communication constraints and\vspace{-0.1cm}
\begin{equation}\label{eq:position_monotone_test}
\mathcal J_T(\mathbf p^{\rm tr}) \le \mathcal J_T(\mathbf p),
\end{equation}
where $\mathbf p^{\rm tr}$ denote the current position vector $\mathbf p$ with its $m$-th block replaced by $\mathbf p_m^{\rm tr}$.

The subarray blocks are updated sequentially using their latest values. A trial point is accepted only if it satisfies the original constraints and \eqref{eq:position_monotone_test}. Otherwise, the current block is retained. This guarantees that the accepted AO objective sequence is non-increasing and bounded below by zero, and therefore converges to a finite value.

\section{Numerical Results and Analysis} \label{Sec:sim_results}
A mono-static ISAC network with $K = 3$ communication users and 1 sensing target is studied where the users and sensing target are dropped within 30-meter ranges from the BS.
These ranges are larger than the Rayleigh distance of each compact subarray but smaller than that of the overall aperture, thereby satisfying the considered hybrid-field condition.
Unless otherwise stated, the BS operates at 28 GHz and employs a $2\times2$ layout of $M = 4$ MSAs. Each subarray is a $4\times4$ UPA with half-wavelength element spacing, resulting in $MN=64$ antennas over a $0.4~\mathrm{m}\times0.4~\mathrm{m}$ overall aperture.

Each communication channel contains one LoS path and two NLoS paths whose powers are $10$ dB and $15$ dB below that of the LoS path, respectively. The common path attenuation is generated by the Friis model. We set $P_{\max}=1$ W, $\sigma_k^2=10^{-11}$ W, $T_s=128$, $I_{\max}=10$ and use the normalized sensing parameters $|\beta_s|=1$ and $\sigma_s^2=1$.
The reported scalar CRB in dB is defined as
$10\log_{10}[\operatorname{tr}(\mathrm{CRB})]$.

\begin{figure}[t]
    \centering
    \includegraphics[width=0.4\textwidth]{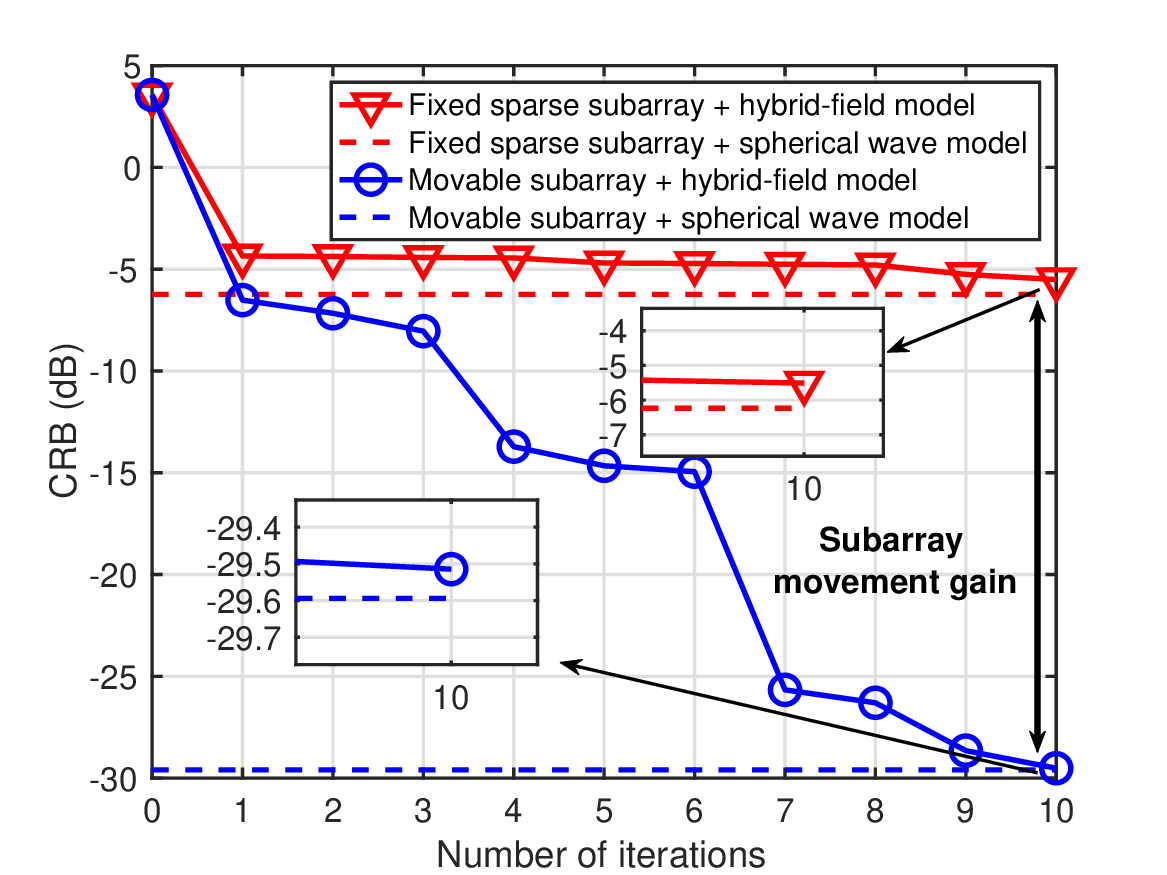}
    \caption{CRB performance comparison under 5 dB SINR constraint.}
    \label{fig:convergence}
   \vspace{-3.5 mm}
\end{figure}

Fig.~\ref{fig:convergence} separates the effects of propagation modeling and subarray mobility. In the fixed sparse-subarray benchmark, each subarray is fixed at the center of each area $\mathcal C_m$, while the beamforming matrix is still optimized using the same rank-one-penalized SDR procedure. 
The solid curves show the CRB evaluated using the proposed hybrid-field model, whereas the dashed lines re-evaluate the same beamforming matrices and subarray positions obtained from the 10th iteration using the spherical-wave model. Therefore, the gap between solid and dashed pair quantifies the modeling mismatch, which is less than 0.75 dB for both fixed and movable subarray schemes. 
Moreover, Fig. \ref{fig:convergence} also demonstrates a gain of 24 dB in CRB given by the MSAs compared to fixed sparse arrays, proving the benefits of utilizing the extra DoF of subarray movement.

\begin{figure}[!t]
    \centering
    \includegraphics[width=0.4\textwidth]{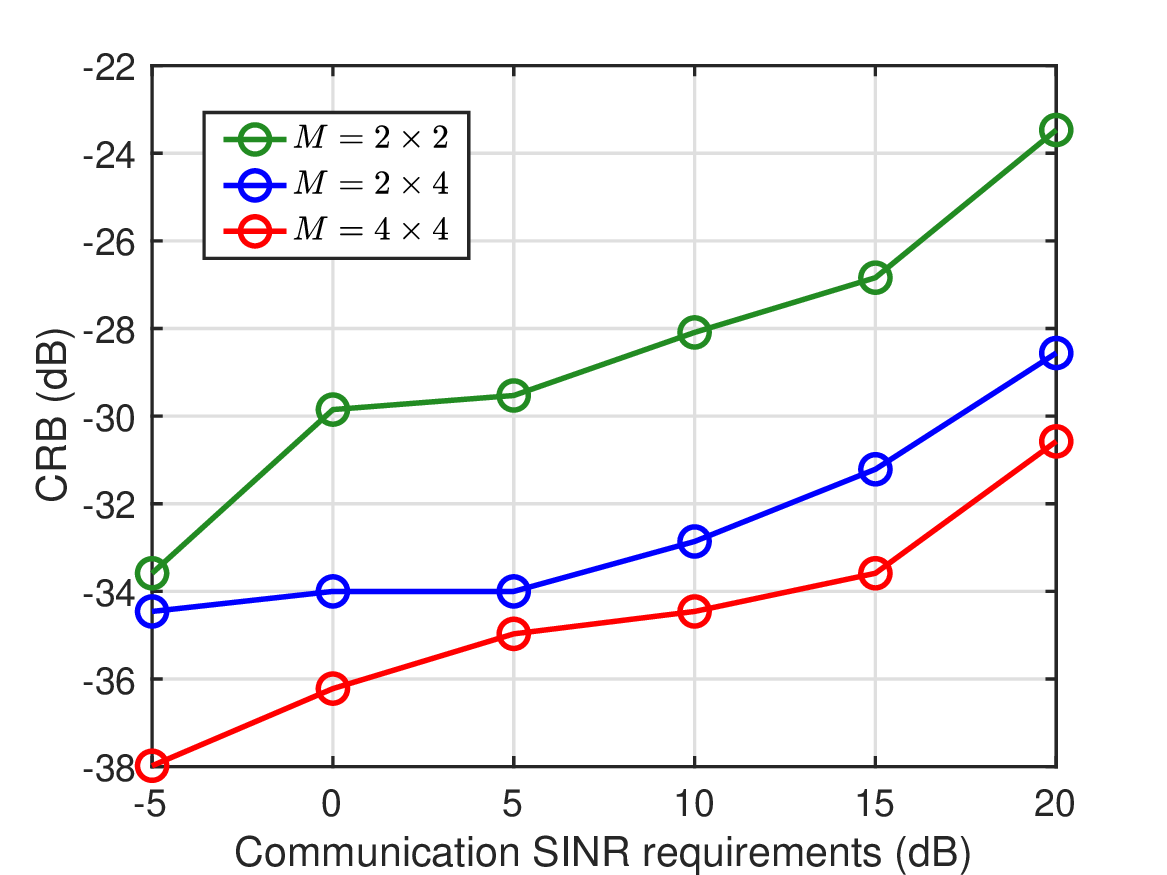}
    \caption{Sensing versus communication performance of MA-aided ISAC systems under different subarray setup.}
    \label{fig:SINR_sweep}
   \vspace{-4 mm}
\end{figure}

Fig.~\ref{fig:SINR_sweep} illustrates the sensing versus communication performance tradeoff. As the required communication SINR increases, the achievable trace CRB increases for all array partitions because more spatial and power resources must be reserved for communications.
Additionally, the experiment is carried out for three different cases where the BS array is divided into 4, 8 or 16 MSAs. For a fixed total aperture and antenna number, partitioning the array into more MSAs will improve system performance. This insight is useful under practical scenarios where the BS array size and the total number of antenna elements are often restricted.

\vspace{-0.2cm}
\section{Conclusion}  \label{Sec:conclusions}
In this letter, we investigate an MSA aided ISAC system using a hybrid near-far field channel model, and propose an AO algorithm to find the suitable subarray positions and beamforming matrices. 
Numerical results show that the proposed hybrid-field model exhibits less than 0.75 dB CRB mismatch compared to the spherical-wave model, which indicates that the proposed hybrid-field model can accurately capture the channel characteristics in the considered regime. 
Simulations also testify the benefits of deploying MSAs compared to fixed subarrays by showing a 24 dB gain in trace CRB.
Tradeoff between communication and sensing performance is also simulated which reveals the benefit of dividing the whole array into more smaller subarrays.

\vspace{-0.2cm}

\bibliographystyle{IEEEtran}
\bibliography{Reference}

\end{document}